\documentclass[conference,a4paper]{IEEEtran}
\IEEEoverridecommandlockouts
\usepackage{cite}
\usepackage{amsmath,amssymb,amsfonts}
\usepackage{algorithmic}
\usepackage{graphicx}
\usepackage{tabularx}
\usepackage{booktabs}
\usepackage{textcomp}
\usepackage{subcaption}
\usepackage{xcolor}
\def\BibTeX{{\rm B\kern-.05em{\sc i\kern-.025em b}\kern-.08em
    T\kern-.1667em\lower.7ex\hbox{E}\kern-.125emX}}

\usepackage{tikz}
    \usetikzlibrary{shapes.arrows}

\usepackage{pgfplots}
\pgfplotsset{compat=1.16}
\usepackage{adjustbox}

\usepackage{gincltex}
\usepgfplotslibrary{fillbetween}
\usepgfplotslibrary{groupplots}
\usetikzlibrary{plotmarks}
\usetikzlibrary{patterns}
\pgfplotsset{
tick label style={font=\footnotesize},
label style={font=\footnotesize},
legend style={font=\footnotesize},
}

\definecolor{violet}{rgb}{0.6,0,0.6}%
\definecolor{orange_D}{rgb}{1,0.3,0}%
\definecolor{cyan}{rgb}{0,0.67,0.64}%
\definecolor{red}{rgb}{0.9,0,0}%
\definecolor{green}{rgb}{0,0.8,0}%
\definecolor{yellow}{rgb}{1,0.8,0}

\def \fwidth{0.9\columnwidth}
\def \fheight {0.45\columnwidth}

\usepackage{glossaries}    
    
\newacronym[plural=MDPs,firstplural=Markov Decision Processes (MDPs)]{mdp}{MDP}{Markov Decision Process}
\newacronym{iot}{IoT}{Internet of Things}
\newacronym{fec}{FEC}{Forward Error Correction}
\newacronym{snr}{SNR}{Signal to Noise Ratio}
\newacronym{aoi}{AoI}{Age of Information}
\newacronym{paoi}{PAoI}{Peak Age of Information}
\newacronym{qaoi}{QAoI}{Age of Information at Query}
\newacronym{pec}{PEC}{Packet Erasure Channel}
\newacronym{pdf}{PDF}{Probability Density Function}
\newacronym{cdf}{CDF}{Cumulative Distribution Function}
\newacronym{pq}{PQ}{Permanent Query}
\newacronym{qapa}{QAPA}{Query Arrival Process Aware}
    
\pgfplotsset{compat=1.16}

\begin{document}

\title{Freshness on Demand: Optimizing Age of Information for the Query Process}

\author{\IEEEauthorblockN{Josefine Holm$^*$, Anders E. Kal{\o}r$^*$, Federico Chiariotti$^*$, Beatriz Soret$^*$, Søren K. Jensen$^+$, \\Torben B. Pedersen$^+$, and Petar Popovski$^*$}
\IEEEauthorblockA{$^*$Department of Electronic Systems, Aalborg University\\
Fredrik Bajers Vej 7C, 9220 Aalborg, Denmark, email: \{jho,aek,fchi,bsa,petarp\}@es.aau.dk\\
$^+$Department of Computer Science, Aalborg University\\
Selma Lagerl{\o}fs Vej 300, 9220 Aalborg, Denmark, email: \{skj,tbp\}@cs.aau.dk}
}

\maketitle

\begin{abstract}
\gls{aoi} has become an important concept in communications, as it allows system designers to measure the freshness of the information available to remote monitoring or control processes. However, its definition tacitly assumed that new information is used at any time, which is not always the case and the instants at which information is collected and used are dependent on a certain query process.  
We propose a model that accounts for the discrete time nature of many monitoring processes, considering a \emph{pull-based communication model} in which the freshness of information is only important when the receiver generates a query. We then define the \gls{qaoi}, a more general metric that fits the pull-based scenario, and show how its optimization can lead to very different choices from traditional push-based \gls{aoi} optimization when using a \gls{pec}.
\end{abstract}

\glsresetall


\section{Introduction}
Over the past few years, the concept of information freshness has received a significant attention in relation to cyber-physical systems that rely on communication of various updates in real time. This has led to the introduction of \emph{\gls{aoi}}~\cite{kosta2017age} as a measure that reflects the freshness at the receiver with respect to the sender, and denotes the difference between the current time and the time when the most recently received update was generated at the sender.

The first works to deal with \gls{aoi} considered simple queuing systems, deriving analytical formulas for information freshness~\cite{kaul2012real}. Followup works addressed \gls{aoi} in specific wireless scenarios with errors~\cite{chen2016error} and retransmissions~\cite{devassy2019reliable}, or basing their analysis on live experiments~\cite{beytur2019measuring}. The addition of more sources in the queuing system leads to an interesting scheduling problem, which aims at finding the packet generation rate that minimizes the age for the whole system~\cite{kadota2019minimizing}. Optimizing the access method and senders' updating policies to minimize \gls{aoi} in complex wireless communication system has been proven to be an NP-hard problem, but heuristics can achieve near-optimal solutions~\cite{sun2017update} by having sources decide whether an update is valuable enough to be sent (i.e., whether it would significantly reduce the \gls{aoi}) \cite{yates2015lazy}. The average \gls{aoi} has been derived in slotted~\cite{yates2017unreliable} and unslotted ALOHA~\cite{yates2020unslotted}, as well as in scheduled access~\cite{talak2018distributed}, and the performance of scheduling policies has been combined with these access methods in~\cite{chen2020rach}.

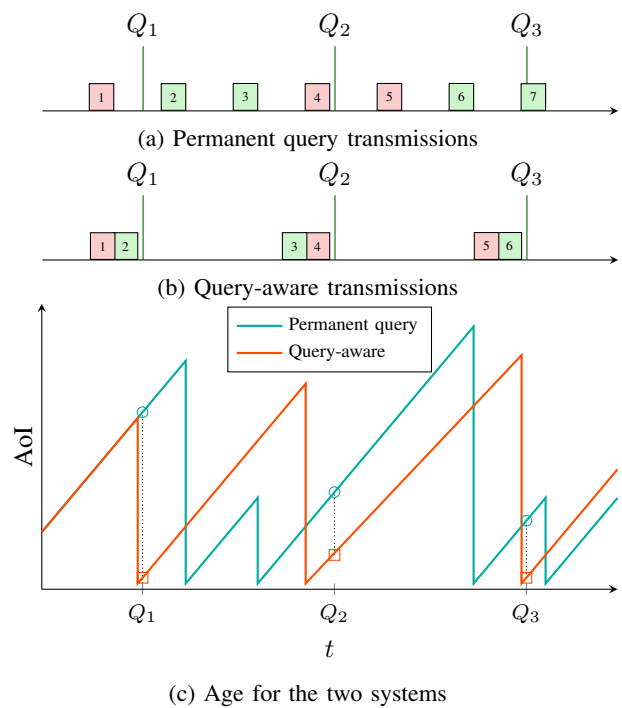
\begin{figure}
	\flushright
	\begin{subfigure}[b]{0.95\columnwidth}
		\flushright
        \begin{tikzpicture}

\begin{axis}[%
width=\fwidth,
height=\fheight/3,
name=plot1,
scale only axis,
ymajorticks=false,
xmajorticks=false,
clip=false,
axis x line=bottom,
axis y line=none,
xmin=0,
xmax=120,
ylabel near ticks,
ymin=0,
ymax=1,
legend style={font=\scriptsize, at={(0.95,0.95)}, anchor=north east, legend cell align=left, align=left, draw}
]

\node at (axis cs:21,0.9)[name=q1] {$Q_1$};
\node at (axis cs:61,0.9)[name=q2] {$Q_2$};
\node at (axis cs:101,0.9)[name=q3] {$Q_3$};

\draw[-,green!50.19607843137255!black] (q1.south) -- (axis cs:21,0);
\draw[-,green!50.19607843137255!black] (q2.south) -- (axis cs:61,0);
\draw[-,green!50.19607843137255!black] (q3.south) -- (axis cs:101,0);
\node at (axis cs:15,0) [draw,anchor=south east,minimum width=0.3cm,minimum height=0.25cm,fill={white!80!red}] {\tiny{1}};
\node at (axis cs:30,0) [draw,anchor=south east,minimum width=0.3cm,minimum height=0.25cm,fill={white!80!green}] {\tiny{2}};
\node at (axis cs:45,0) [draw,anchor=south east,minimum width=0.3cm,minimum height=0.25cm,fill={white!80!green}] {\tiny{3}};
\node at (axis cs:60,0) [draw,anchor=south east,minimum width=0.3cm,minimum height=0.25cm,fill={white!80!red}] {\tiny{4}};
\node at (axis cs:75,0) [draw,anchor=south east,minimum width=0.3cm,minimum height=0.25cm,fill={white!80!red}] {\tiny{5}};
\node at (axis cs:90,0) [draw,anchor=south east,minimum width=0.3cm,minimum height=0.25cm,fill={white!80!green}] {\tiny{6}};
\node at (axis cs:105,0) [draw,anchor=south east,minimum width=0.3cm,minimum height=0.25cm,fill={white!80!green}] {\tiny{7}};
\end{axis}
\end{tikzpicture}
        \caption{Permanent query transmissions}
        \label{fig:qp}
    \end{subfigure}	
    \begin{subfigure}[b]{0.95\columnwidth}
    	\flushright
        \begin{tikzpicture}

\begin{axis}[%
width=\fwidth,
height=\fheight/3,
name=plot1,
scale only axis,
ymajorticks=false,
xmajorticks=false,
clip=false,
axis x line=bottom,
axis y line=none,
xmin=0,
xmax=120,
ylabel near ticks,
ymin=0,
ymax=1,
legend style={font=\scriptsize, at={(0.95,0.95)}, anchor=north east, legend cell align=left, align=left, draw}
]

\node at (axis cs:21,0.9)[name=q1] {$Q_1$};
\node at (axis cs:61,0.9)[name=q2] {$Q_2$};
\node at (axis cs:101,0.9)[name=q3] {$Q_3$};

\draw[-,green!50.19607843137255!black] (q1.south) -- (axis cs:21,0) ;
\draw[-,green!50.19607843137255!black] (q2.south) -- (axis cs:61,0);
\draw[-,green!50.19607843137255!black] (q3.south) -- (axis cs:101,0);
\node at (axis cs:20,0) [draw,anchor=south east,minimum width=0.3cm,minimum height=0.25cm,fill={white!80!green},name=p2] {\tiny{2}};
\node [draw,left of=p2,node distance=0.3cm,minimum width=0.3cm,fill={white!80!red},minimum height=0.25cm,name=p1] {\tiny{1}};
\node at (axis cs:60,0) [draw,anchor=south east,minimum width=0.3cm,minimum height=0.25cm,fill={white!80!red},name=p4] {\tiny{4}};
\node [draw,left of=p4,node distance=0.3cm,minimum width=0.3cm,fill={white!80!green},minimum height=0.25cm,name=p3] {\tiny{3}};
\node at (axis cs:100,0) [draw,anchor=south east,minimum width=0.3cm,minimum height=0.25cm,fill={white!80!green},name=p6] {\tiny{6}};
\node [draw,left of=p6,node distance=0.3cm,minimum width=0.3cm,fill={white!80!red},minimum height=0.3cm,name=p5] {\tiny{5}};

\end{axis}
\end{tikzpicture}
        \caption{Query-aware transmissions}
        \label{fig:qapa}
    \end{subfigure}
	\begin{subfigure}[b]{0.95\columnwidth}
		\flushright
        \begin{tikzpicture}

\begin{axis}[%
width=\fwidth,
height=\fheight,
name=plot1,
scale only axis,
xtick={21,61,101},
xticklabels={$Q_1$,$Q_2$,$Q_3$},
ymajorticks=false,
clip=false,
axis x line=bottom,
axis y line=left,
xlabel near ticks,
xlabel={$t$},
xmin=0,
xmax=120,
ylabel near ticks,
ymin=0,
ymax=50,
ylabel={AoI},
legend style={font=\scriptsize, at={(0.5,0.99)}, anchor=north, legend cell align=left, align=left, draw}
]
\draw[-,densely dotted] (axis cs:21,1) -- (axis cs:21,31) ;
\draw[-,densely dotted] (axis cs:61,6) -- (axis cs:61,17) ;
\draw[-,densely dotted] (axis cs:101,2) -- (axis cs:101,12) ;

\addplot [color=cyan,line width=0.8pt]
  table[row sep=crcr]{
0	10\\
30  40\\
30 1\\
45 16\\
45 1\\
90 46\\
90 1\\
105 16\\
105 1\\
120 16\\
};
\addlegendentry{Permanent query}

\addplot [color=cyan,only marks,  mark=o, mark size=2, forget plot]
table[row sep=crcr]{
21 31\\
61 17\\
101 12\\
};

\addplot [color=orange_D, line width=0.8pt]
  table[row sep=crcr]{
0	10\\
20 30\\
20 1\\
55 36\\
55 1\\
100 41\\
100 1\\
120 21\\
};
\addlegendentry{Query-aware}

\addplot [color=orange_D,only marks,  mark=square, mark size=2, forget plot]
table[row sep=crcr]{
21 2\\
61 6\\
101 2\\
};

\end{axis}
\end{tikzpicture}
        \caption{Age for the two systems}
        \label{fig:age}
	\end{subfigure}
 \caption{Example of the difference between a system assuming a permanent query and one that is aware of the query arrival process. The same packets are lost (depicted in red) in both systems, and the markers indicate the age at the query arrival instants.}
 \label{fig:toy_example}
\end{figure}

However, the tacit assumption behind \gls{aoi}, regardless of the system for which it is computed, has been that the receiver is interested in having fresh information \emph{at any time}. In other words, this assumption works with a \emph{push-based communication}, in which a hypothetical application residing at the receiver has a \emph{permanent query} to the updates that arrive at the receiver. The motivation for this paper starts by questioning this underlying assumption and generalizes the models
used in \gls{aoi} by considering the timing of the query process. This makes the model into \emph{pull-based communication}, where the query instants can guide the communication strategy for the sensor updates. 

The impact of the query-driven, pull-based communication model becomes immediately obvious with the (over)simplified example in Fig.~\ref{fig:toy_example}. The time is slotted and each update packet, labeled $1,2, \ldots 7$, takes one slot. Each update is generated immediately prior to the transmission.
The queries $Q_1, Q_2, Q_3, \ldots$ arrive periodically, every $7$-th slot. Furthermore, as an energy constraint, it is assumed that the sender can transmit on average one packet every $3$ slots. Fig.~\ref{fig:qp} shows the case in which the sender is oblivious to the query arrival process and distributes the transmissions evenly in time. Another strategy could be, in each slot, to decide to transmit with probability $1/3$ or stay silent otherwise; the important point is that this decision is made independently from the query process. Fig.~\ref{fig:qapa} shows query-driven communication: the sender knows the query instants and optimizes the transmissions with respect to the timing of the query process, i.e., sends just before the query instants. In both cases the (red) packets $1,4,5$ are lost due to transmission errors. Fig.~\ref{fig:age} shows that the query-driven strategy is more likely to provide updates that are fresh when a query arrives, although its average \gls{aoi} is worse at the instants in which there is no query.

Despite the deceptively simple insight offered by the example from Fig.~\ref{fig:toy_example}, the introduction of query-driven communication strategies does have a practical significance and 
introduces novel and interesting problems, as this paper shows. In fact, the assumption of a permanent query is relatively uncommon in the network control literature~\cite{zhang2006communication}, which often uses periodic discrete time systems that poll the state of the monitored process at predefined intervals. Most network control system are asynchronous, using different sampling strategies that depend on the reliability of the connection and on the monitored process~\cite{zhang2017analysis}. We define a \emph{query arrival process} and consider the optimization of the communication process with respect to that arrival process. Furthermore, we define a \gls{qaoi} metric which reflects the freshness in the instants when the receiver actually needs the data: having fresh data when the monitoring process is not asking for it does not provide any benefits to the system, as the information will not be used. Our model is also relevant for duty cycle-based applications, in which the sleeping pattern of the sensors are synchronized with the monitoring process.

This paper introduces models to  analyze the difference in the communication strategies that should be used when the query arrival process is taken into account compared to the treatment of \gls{aoi} in the context of a permanent query. In this initial work, we derive a \gls{mdp} model for the problem with periodic queries and an erasure channel, and show that an optimization aimed at \gls{qaoi} can significantly improve the perceived freshness with respect to classical models.

The remainder of the paper is organized as follows. We define the system model and the concept of \gls{qaoi} in Sec.~\ref{sec:system}, and we formalize it as an \gls{mdp} in Sec.~\ref{sec:mdp}. The setting and results of our simulations are described in Sec.~\ref{sec:results}, and Sec.~\ref{sec:concl} concludes the paper and presents some possible avenues of future work.

\section{System model}\label{sec:system}

We consider a scenario in which a wireless sensor generates updates at will and transmits them to an edge node over a wireless channel. The edge node receives queries from a server about the state of the sensor, e.g. as part of a monitoring or control process. The objective of this work is to maximize the freshness of the information used in the query responses while considering that the sensor is energy-constrained and needs to limit the number of transmissions to the edge node to prolong its lifetime.

\subsection{Age of Information at Query}\label{ssec:qaoi}

We consider a time-slotted system indexed by $t=1,2,\ldots$, and denote the time instances at which updates are successfully delivered to the edge node by $t_{u,1},t_{u,2},\ldots$. Following the common definition of \gls{aoi} considered in the literature, e.g. \cite{kaul2012real, kadota2019minimizing} we denote the \gls{aoi} in time slot $t$ by $\Delta(t)$, and define it as the difference between $t$ and the time at which the last successfully received packet was generated:
\begin{equation}
    \Delta(t)=t-\max_{i:t_{u,i}\le t} t_{u,i}.
\end{equation}
We will assume that $t_{u,1}=0$ so that $\Delta(t)$ is well defined. An alternative, but equivalent definition can be obtained by letting $u_t=1$ if a new update is received at the edge note in time slot $t$, otherwise $u_t=0$:
\begin{equation}
    \Delta(t)=\begin{cases}
    \Delta(t-1) +1 & \text{if } u_t=0\\
    1 & \text{if } u_t=1
    \end{cases}
\end{equation}
where $\Delta(0)=0$.

Most work considers the problem of maximizing the long-term average of $\Delta(t)$. However, this is only one possibility in real monitoring and control systems: discrete-time systems involve queries in which the monitoring process samples the available information. To capture such applications, we introduce the \gls{qaoi} metric, which generalizes \gls{aoi} by sampling $\Delta(t)$ according to an arbitrary querying process, thereby considering only the instants at which a query arrives. We denote the query arrival times at the edge node by $t_{q,1},t_{q,2},\ldots$, and define the overall objective as minimizing the long-term expected \gls{qaoi} defined as
\begin{equation}
    \tau_{\infty}=\lim_{t\to\infty} \mathbb{E}\left[\sum_{i: t_{q,i}\le t} \Delta(t_{q,i})\right].
\end{equation}
Although the query process may in general follow any random process, in this initial paper we limit the focus to the case in which the exact query instants are know in advance to the edge node and the sensor. This is for instance the case when the queries are periodic, or if the server repeatedly announces its next query instant.

\subsection{Models for Communication and Query Arrivals}
We assume that each update has a fixed size and is transmitted over a \gls{pec} with erasure probability $\epsilon$. For simplicity's sake, in the following we refer to the success probability $p_s=1-\epsilon$. Packets are instantaneously acknowledged by the receiver, so the sensor knows if a packet was erased or correctly received.

To model the energy-constrained nature of the node, we use a \emph{leaky bucket} model, as commonly done in the literature~\cite{raghunathan2004energy}: we consider a bucket of tokens, which is replenished by a process which can generate tokens independently at each step with probability $\mu_b$. The node can only transmit a packet if there are tokens in the bucket, and each transmission consumes one token. This model can fit an energy gathering node, as well as a general power consumption constraint on a battery-powered node, which should limit its number of transmissions in order to prolong its lifetime.

In this work, we assume the simplest possible query arrival process, with periodic queries every $T_q$ steps. We assume that the sensor and receiver are synchronized, i.e., the sensor knows when the next query will come. While simple, this assumption is often realistic, as discrete time monitoring processes are often designed with a constant time step. 

The model can be easily extended to more complex query arrival processes, and the process statistics can even be learned implicitly as part of the optimal strategy, as long as it is consistent. If we follow the definitions from Sec.~\ref{ssec:qaoi}, the strategies to minimize \gls{aoi} and \gls{qaoi} coincide in the memoryless case in which the query arrival process is Poisson or when the query arrival process is much faster than the sensor, i.e., when there is a query in each time slot.

\section{MDP formulation and problem solution}\label{sec:mdp}
In the following, we will model the two communication scenarios described in the next paragraph 
as \glspl{mdp}, which we will then proceed to solve. An \gls{mdp} is defined by a state space 
$\mathcal{S}$, an action space $\mathcal{A}$, a set of transition probabilities $p_a(s,s')=P(s_{t+1}=s'|a_t=a,s_t=s)$,  and an instantaneous reward function $r(s,a,s')$, which represents the immediate reward when taking action $a$ and transitioning from state $s$ to state $s'$. 
The model can be used to represent two different systems: a \gls{pq} system, which minimizes the traditional \gls{aoi}, and a \gls{qapa} system, which minimizes the \gls{qaoi}, only caring about the instants when a query arrives. These two systems can use the same state and action spaces, and only differ in the reward function that they use.

Decisions are made at every slot, as the sensor can either keep silent or send a packet. Consequently, the action space is $\mathcal{A}=\{0,1\}$. As the aim of the \gls{qapa} agent is to minimize the \gls{qaoi}, the state should include the current age $\Delta(t)$, as well as the number of slots $\sigma(t)$ until the next query. Additionally, the agent should know the number of available tokens, $b(t)$, as it will influence its decision whether to transmit. If the number of tokens is 0, the sensor is blocked from transmitting until a token is generated. The state space can then be defined as $\mathcal{S}=\mathbb{N}\times\{0,\ldots,T_q-1\}\times \mathbb{N}$, where $\mathbb{N}$ indicates the set of strictly positive integers. A state $s_t$ is given by the tuple $(\Delta(t),\sigma(t),b(t))$. Each element in the state tuple evolves independently between time step, so in the following we describe the state dynamics one by one.

The \gls{aoi} increases by one between each slot unless the node decides to transmit and the packet is successfully received, with probability $p_s$, in which case the \gls{aoi} is reduced to one in the subsequent slot. The non-zero transition probabilities are thus described by
\begin{align}
    &\Pr(\Delta(t+1)=\delta|\Delta(t),a_t) = \begin{cases} a_tp_s &\text{if }\delta=1;\\
    1-a_tp_s &\text{if }\delta=\Delta(t)+1;\\
    0 &\text{otherwise,}
    \end{cases}
\end{align}
where $a_t$ is the action at time $t$, which equals zero if the sensor is silent and one if it transmits.
The time until the next query $\sigma(t)$, is deterministic and independent of the action, and decreases by one until it reaches zero, at which point it is reset to $T_q$. Assuming that the first query happens at time $t=T_q$, the value of $\sigma(t)$ can be written
\begin{equation}
    \sigma(t) = T_q-t \pmod{T_q}.
\end{equation}
Finally, the number of tokens in the next slot depends on whether a new token is generated and whether the sensor transmits, in which case, it uses one token. The transition probability from $b(t)$ to $b(t+1)$ is:
\begin{equation}
p(b(t+1)=b+i|b(t),a_t)=\begin{cases}
 \mu_b&\text{if }i=1-a_t;\\
 1-\mu_b&\text{if }i=-a_t;\\
 0 &\text{otherwise.}
\end{cases}
\end{equation}
We define two cost functions; one for the \gls{pq} system, which does not depend on the query instant and will be used as baseline, and one for the \gls{qapa} system, in which the cost is only considered when a query arrives. In the baseline \gls{pq} model, the cost is given by the AoI in any slot:
\begin{equation}\label{eq:cost_AoI}
    c_{\text{PQ}}(s_t,a_t,s_{t+1})=\Delta(t+1).
\end{equation}
However, in the \gls{qapa} system, the cost is the AoI when a query arrives:
\begin{equation}\label{eq:cost_QAoI}
    c_{\text{QAPA}}(s_t,a_t,s_{t+1})=\begin{cases}
    \Delta(t+1) &\text{if }\sigma(t+1)= 0;\\
    0 &\text{otherwise.}
    \end{cases}
\end{equation}
In both cases, the objective is to find a policy $\pi^*$ that minimizes the long-term cost. In this initial work, we limit ourselves to consider the discounted case, which benefits from strong convergence guarantees, and defer the case with undiscounted costs to future work. Specifically, we solve:
\begin{equation}
\pi^*=\arg\min_{\pi} \mathbb{E}\left[\sum_{t=0}^\infty \lambda^t c(t)|\pi\right],
\end{equation}
where $\lambda< 1$ is the discount factor.

We can now proceed to solve the \gls{mdp} for the two systems we have defined using policy iteration, as described in \cite[Ch. 4]{sutton2018reinforcement}. In order to apply the algorithm, we need to truncate the problem to a finite \gls{mdp}. We do so by defining a maximum age $\Delta_{\max}$ and a token bucket size $B$: once the age or the number of tokens in the bucket reach the maximum, they cannot increase further. As long as the maximum values are sufficiently large, they are not reached during normal operation and this simplification does not affect the optimal policies or their performance.

The policy iteration algorithm has two steps: 1) policy evaluation and 2) policy improvement which are repeated until convergence. To solve the proposed problem we initialize the policy with zeros i.e. the policy where we never send any updates, and the value to be larger than we expect from a reasonable policy. 
\begin{enumerate}
    \item The policy is evaluated using \begin{equation}
        v_\pi(s)=\sum_{s'}p(s',c\vert s,a)\left(c+\lambda v_\pi(s')\right).
    \end{equation}
     for all $s$, where $s$ is the current state, $s'$ is the new state, $a$ is the action, and $c$ is the cost from either \eqref{eq:cost_AoI} or \eqref{eq:cost_QAoI}.
    \item The policy is improved by evaluating \begin{equation}
        q_\pi(s,a)=\sum_{s'}p(s',c\vert s,a)\left(c+\lambda v_\pi(s')\right)
    \end{equation}
    for all $a$. If $q_\pi(s,a)>v_\pi(s)$ we substitute $a$ into the policy. This is repeated for all $s$.
\end{enumerate}
Policy iteration is guaranteed to converge to the optimal policy~\cite{howard1960dynamic} in finite-state \glspl{mdp} with finite reward. As mentioned above, we truncated the age and token bucket size to make the \gls{mdp} finite, so the conditions to use the algorithm apply.

\section{Numerical results}\label{sec:results}
This section presents Monte Carlo evaluations of the policies obtained using the \gls{mdp} described in Section~\ref{sec:mdp}. Although, the methods in Section~\ref{sec:mdp} can be applied to any query process, throughout the evaluation we will consider queries that occur periodically, at a fixed time interval $T_q$. Furthermore, we truncate the \gls{mdp} at a maximum age of $\Delta_{\text{max}}=100\times T_q$ and a maximum token bucket size of $B=10$, and we use a discount factor $\gamma=0.75$.
We use the term \gls{aoi} to refer to the age at any time and \gls{qaoi} for the age sampled at the query instants.

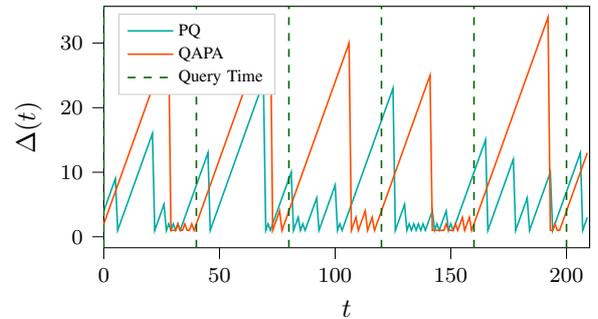
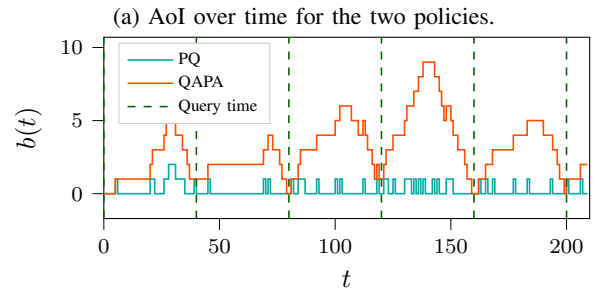
\begin{figure}
    \centering
    	\begin{subfigure}[b]{.9\columnwidth}
	    \centering
    % This file was created by matplotlib2tikz v0.7.5.
\begin{tikzpicture}

\definecolor{color0}{rgb}{0.12156862745098,0.466666666666667,0.705882352941177}
\definecolor{color1}{rgb}{1,0.498039215686275,0.0549019607843137}

\begin{axis}[
height=0.6\linewidth,
width=\linewidth,
legend cell align={left},
legend style={at={(0.03,0.97)}, anchor=north west, draw=white!80.0!black,nodes={scale=0.8, transform shape}},
tick align=outside,
tick pos=left,
x grid style={white!69.01960784313725!black},
xlabel={$t$},
xmin=0, xmax=210,
xtick style={color=black},
y grid style={white!69.01960784313725!black},
ylabel={$\Delta(t)$},
ymin=-1.7, ymax=35.7,
ytick style={color=black}
]
\addplot [semithick, color=cyan]
table {%
0 4
1 5
2 6
3 7
4 8
5 9
6 1
7 2
8 3
9 4
10 5
11 6
12 7
13 8
14 9
15 10
16 11
17 12
18 13
19 14
20 15
21 16
22 1
23 2
24 3
25 4
26 5
27 1
28 2
29 1
30 2
31 1
32 2
33 1
34 2
35 3
36 4
37 5
38 6
39 7
40 8
41 9
42 10
43 11
44 12
45 13
46 1
47 2
48 3
49 4
50 5
51 6
52 7
53 8
54 9
55 10
56 11
57 12
58 13
59 14
60 15
61 16
62 17
63 18
64 19
65 20
66 21
67 22
68 23
69 24
70 1
71 2
72 1
73 2
74 3
75 4
76 5
77 6
78 7
79 8
80 9
81 10
82 1
83 2
84 3
85 1
86 2
87 1
88 2
89 3
90 4
91 5
92 6
93 1
94 2
95 3
96 4
97 5
98 6
99 7
100 8
101 1
102 2
103 1
104 2
105 3
106 4
107 5
108 6
109 7
110 8
111 9
112 10
113 11
114 12
115 13
116 14
117 15
118 16
119 17
120 18
121 19
122 20
123 21
124 22
125 23
126 1
127 2
128 3
129 4
130 5
131 1
132 2
133 1
134 2
135 1
136 2
137 1
138 2
139 1
140 2
141 3
142 4
143 1
144 2
145 1
146 2
147 3
148 4
149 1
150 2
151 1
152 2
153 3
154 4
155 5
156 6
157 7
158 8
159 9
160 10
161 11
162 12
163 13
164 14
165 15
166 1
167 2
168 3
169 4
170 5
171 6
172 7
173 8
174 9
175 10
176 11
177 12
178 1
179 2
180 3
181 4
182 5
183 6
184 1
185 2
186 3
187 4
188 5
189 6
190 7
191 8
192 9
193 10
194 1
195 2
196 3
197 4
198 5
199 6
200 7
201 8
202 9
203 10
204 11
205 12
206 13
207 1
208 2
209 3
};
\addlegendentry{PQ}
\addplot [semithick, orange_D]
table {%
0 2
1 3
2 4
3 5
4 6
5 7
6 8
7 9
8 10
9 11
10 12
11 13
12 14
13 15
14 16
15 17
16 18
17 19
18 20
19 21
20 22
21 23
22 24
23 25
24 26
25 27
26 28
27 29
28 30
29 1
30 1
31 1
32 2
33 1
34 1
35 2
36 1
37 1
38 2
39 1
40 2
41 3
42 4
43 5
44 6
45 7
46 8
47 9
48 10
49 11
50 12
51 13
52 14
53 15
54 16
55 17
56 18
57 19
58 20
59 21
60 22
61 23
62 24
63 25
64 26
65 27
66 28
67 29
68 30
69 31
70 32
71 33
72 34
73 1
74 2
75 3
76 4
77 1
78 2
79 3
80 4
81 5
82 6
83 7
84 8
85 9
86 10
87 11
88 12
89 13
90 14
91 15
92 16
93 17
94 18
95 19
96 20
97 21
98 22
99 23
100 24
101 25
102 26
103 27
104 28
105 29
106 30
107 1
108 2
109 3
110 1
111 2
112 3
113 4
114 1
115 2
116 3
117 1
118 2
119 3
120 4
121 5
122 6
123 7
124 8
125 9
126 10
127 11
128 12
129 13
130 14
131 15
132 16
133 17
134 18
135 19
136 20
137 21
138 22
139 23
140 24
141 25
142 1
143 1
144 1
145 1
146 1
147 1
148 2
149 1
150 1
151 1
152 2
153 1
154 2
155 3
156 1
157 2
158 3
159 1
160 2
161 3
162 4
163 5
164 6
165 7
166 8
167 9
168 10
169 11
170 12
171 13
172 14
173 15
174 16
175 17
176 18
177 19
178 20
179 21
180 22
181 23
182 24
183 25
184 26
185 27
186 28
187 29
188 30
189 31
190 32
191 33
192 34
193 1
194 1
195 2
196 1
197 1
198 2
199 3
200 4
201 5
202 6
203 7
204 8
205 9
206 10
207 11
208 12
209 13
};
\addlegendentry{QAPA}
\addplot [semithick, green!50.19607843137255!black,dashed]
table {%
0 -1.7
0 35.7
};
\addlegendentry{Query Time}
\addplot [semithick, green!50.19607843137255!black, dashed, forget plot]
table {%
40 -1.7
40 35.7
};
\addplot [semithick, green!50.19607843137255!black, dashed, forget plot]
table {%
80 -1.7
80 35.7
};
\addplot [semithick, green!50.19607843137255!black, dashed, forget plot]
table {%
120 -1.7
120 35.7
};
\addplot [semithick, green!50.19607843137255!black, dashed, forget plot]
table {%
160 -1.7
160 35.7
};
\addplot [semithick, green!50.19607843137255!black, dashed, forget plot]
table {%
200 -1.7
200 35.7
};
\addplot [semithick, green!50.19607843137255!black, dashed, forget plot]
table {%
240 -1.7
240 35.7
};
\addplot [semithick, green!50.19607843137255!black, dashed, forget plot]
table {%
280 -1.7
280 35.7
};
\addplot [semithick, green!50.19607843137255!black, dashed, forget plot]
table {%
320 -1.7
320 35.7
};
\addplot [semithick, green!50.19607843137255!black, dashed, forget plot]
table {%
360 -1.7
360 35.7
};
\addplot [semithick, green!50.19607843137255!black, dashed, forget plot]
table {%
400 -1.7
400 35.7
};
\addplot [semithick, green!50.19607843137255!black, dashed, forget plot]
table {%
440 -1.7
440 35.7
};
\addplot [semithick, green!50.19607843137255!black, dashed, forget plot]
table {%
480 -1.7
480 35.7
};
\addplot [semithick, green!50.19607843137255!black, dashed, forget plot]
table {%
520 -1.7
520 35.7
};
\addplot [semithick, green!50.19607843137255!black, dashed, forget plot]
table {%
560 -1.7
560 35.7
};
\addplot [semithick, green!50.19607843137255!black, dashed, forget plot]
table {%
600 -1.7
600 35.7
};
\addplot [semithick, green!50.19607843137255!black, dashed, forget plot]
table {%
640 -1.7
640 35.7
};
\addplot [semithick, green!50.19607843137255!black, dashed, forget plot]
table {%
680 -1.7
680 35.7
};
\addplot [semithick, green!50.19607843137255!black, dashed, forget plot]
table {%
720 -1.7
720 35.7
};
\addplot [semithick, green!50.19607843137255!black, dashed, forget plot]
table {%
760 -1.7
760 35.7
};
\addplot [semithick, green!50.19607843137255!black, dashed, forget plot]
table {%
800 -1.7
800 35.7
};
\addplot [semithick, green!50.19607843137255!black, dashed, forget plot]
table {%
840 -1.7
840 35.7
};
\addplot [semithick, green!50.19607843137255!black, dashed, forget plot]
table {%
880 -1.7
880 35.7
};
\addplot [semithick, green!50.19607843137255!black, dashed, forget plot]
table {%
920 -1.7
920 35.7
};
\addplot [semithick, green!50.19607843137255!black, dashed, forget plot]
table {%
960 -1.7
960 35.7
};
\end{axis}
\end{tikzpicture}
        \caption{\gls{aoi} over time for the two policies.}
        \label{fig:sawtooth}
    \end{subfigure}	
    	\begin{subfigure}[b]{.9\columnwidth}
	    \centering
        % This file was created by matplotlib2tikz v0.7.5.
\begin{tikzpicture}

\begin{axis}[
height=0.5\linewidth,
width=\linewidth,
legend cell align={left},
legend style={at={(0.03,0.97)}, anchor=north west, draw=white!80.0!black,nodes={scale=0.8, transform shape}},
tick align=outside,
tick pos=left,
x grid style={white!69.01960784313725!black},
xlabel={$t$},
xmin=0, xmax=210,
xtick style={color=black},
y grid style={white!69.01960784313725!black},
ylabel={$b(t)$},
ymin=-1.7, ymax=10.7,
ytick style={color=black}
]
\addplot [semithick, cyan,const plot]
table {%
0 0
1 0
2 0
3 0
4 0
5 1
6 0
7 0
8 0
9 0
10 0
11 0
12 0
13 0
14 0
15 0
16 0
17 0
18 0
19 0
20 1
21 1
22 0
23 0
24 0
25 0
26 1
27 1
28 2
29 2
30 2
31 1
32 1
33 1
34 1
35 0
36 0
37 0
38 0
39 1
40 0
41 0
42 0
43 0
44 0
45 1
46 0
47 0
48 0
49 0
50 0
51 0
52 0
53 0
54 0
55 0
56 0
57 0
58 0
59 0
60 0
61 0
62 0
63 0
64 0
65 0
66 0
67 0
68 0
69 1
70 0
71 1
72 0
73 0
74 0
75 0
76 0
77 0
78 0
79 0
80 0
81 1
82 0
83 0
84 1
85 1
86 1
87 0
88 0
89 0
90 0
91 0
92 1
93 0
94 0
95 0
96 0
97 0
98 0
99 0
100 1
101 0
102 1
103 0
104 0
105 0
106 0
107 0
108 0
109 0
110 0
111 0
112 1
113 0
114 0
115 0
116 0
117 0
118 1
119 0
120 0
121 1
122 1
123 0
124 0
125 1
126 0
127 0
128 0
129 0
130 1
131 1
132 1
133 0
134 1
135 0
136 1
137 0
138 1
139 0
140 0
141 0
142 1
143 0
144 1
145 0
146 0
147 0
148 1
149 1
150 1
151 0
152 0
153 0
154 0
155 0
156 0
157 0
158 0
159 0
160 0
161 0
162 1
163 0
164 0
165 1
166 0
167 0
168 1
169 0
170 0
171 0
172 0
173 0
174 0
175 0
176 0
177 1
178 0
179 0
180 0
181 0
182 0
183 1
184 0
185 0
186 0
187 0
188 0
189 0
190 0
191 0
192 0
193 1
194 0
195 0
196 0
197 0
198 0
199 0
200 1
201 0
202 0
203 0
204 0
205 0
206 1
207 0
208 0
209 0
};
\addlegendentry{PQ}
\addplot [semithick, orange_D,const plot]
table {%
0 0
1 0
2 0
3 0
4 0
5 1
6 1
7 1
8 1
9 1
10 1
11 1
12 1
13 1
14 1
15 1
16 1
17 1
18 1
19 1
20 2
21 3
22 3
23 3
24 3
25 3
26 4
27 5
28 6
29 6
30 5
31 4
32 4
33 4
34 3
35 3
36 2
37 1
38 1
39 1
40 1
41 1
42 1
43 1
44 1
45 2
46 2
47 2
48 2
49 2
50 2
51 2
52 2
53 2
54 2
55 2
56 2
57 2
58 2
59 2
60 2
61 2
62 2
63 2
64 2
65 2
66 2
67 2
68 2
69 3
70 3
71 4
72 4
73 3
74 3
75 3
76 2
77 1
78 1
79 0
80 0
81 1
82 1
83 1
84 2
85 3
86 3
87 3
88 3
89 3
90 3
91 3
92 4
93 4
94 4
95 4
96 4
97 4
98 4
99 4
100 5
101 5
102 6
103 6
104 6
105 6
106 6
107 5
108 5
109 5
110 4
111 4
112 5
113 4
114 3
115 3
116 2
117 1
118 2
119 1
120 1
121 2
122 3
123 3
124 3
125 4
126 4
127 4
128 4
129 4
130 5
131 6
132 6
133 6
134 7
135 7
136 8
137 8
138 9
139 9
140 9
141 9
142 9
143 8
144 8
145 7
146 6
147 5
148 6
149 6
150 5
151 4
152 4
153 3
154 3
155 3
156 2
157 1
158 1
159 0
160 0
161 0
162 1
163 1
164 1
165 2
166 2
167 2
168 3
169 3
170 3
171 3
172 3
173 3
174 3
175 3
176 3
177 4
178 4
179 4
180 4
181 4
182 4
183 5
184 5
185 5
186 5
187 5
188 5
189 5
190 4
191 4
192 4
193 4
194 3
195 3
196 2
197 1
198 1
199 0
200 1
201 1
202 1
203 1
204 1
205 1
206 2
207 2
208 2
209 2
};
\addlegendentry{QAPA}
\addplot [semithick, green!50.19607843137255!black,dashed]
table {%
0 -1.7
0 35.7
};
\addlegendentry{Query time}
\addplot [semithick, green!50.19607843137255!black, dashed, forget plot]
table {%
40 -1.7
40 35.7
};
\addplot [semithick, green!50.19607843137255!black, dashed, forget plot]
table {%
80 -1.7
80 35.7
};
\addplot [semithick, green!50.19607843137255!black, dashed, forget plot]
table {%
120 -1.7
120 35.7
};
\addplot [semithick, green!50.19607843137255!black, dashed, forget plot]
table {%
160 -1.7
160 35.7
};
\addplot [semithick, green!50.19607843137255!black, dashed, forget plot]
table {%
200 -1.7
200 35.7
};
\addplot [semithick, green!50.19607843137255!black, dashed, forget plot]
table {%
240 -1.7
240 35.7
};
\addplot [semithick, green!50.19607843137255!black, dashed, forget plot]
table {%
280 -1.7
280 35.7
};
\addplot [semithick, green!50.19607843137255!black, dashed, forget plot]
table {%
320 -1.7
320 35.7
};
\addplot [semithick, green!50.19607843137255!black, dashed, forget plot]
table {%
360 -1.7
360 35.7
};
\addplot [semithick, green!50.19607843137255!black, dashed, forget plot]
table {%
400 -1.7
400 35.7
};
\addplot [semithick, green!50.19607843137255!black, dashed, forget plot]
table {%
440 -1.7
440 35.7
};
\addplot [semithick, green!50.19607843137255!black, dashed, forget plot]
table {%
480 -1.7
480 35.7
};
\addplot [semithick, green!50.19607843137255!black, dashed, forget plot]
table {%
520 -1.7
520 35.7
};
\addplot [semithick, green!50.19607843137255!black, dashed, forget plot]
table {%
560 -1.7
560 35.7
};
\addplot [semithick, green!50.19607843137255!black, dashed, forget plot]
table {%
600 -1.7
600 35.7
};
\addplot [semithick, green!50.19607843137255!black, dashed, forget plot]
table {%
640 -1.7
640 35.7
};
\addplot [semithick, green!50.19607843137255!black, dashed, forget plot]
table {%
680 -1.7
680 35.7
};
\addplot [semithick, green!50.19607843137255!black, dashed, forget plot]
table {%
720 -1.7
720 35.7
};
\addplot [semithick, green!50.19607843137255!black, dashed, forget plot]
table {%
760 -1.7
760 35.7
};
\addplot [semithick, green!50.19607843137255!black, dashed, forget plot]
table {%
800 -1.7
800 35.7
};
\addplot [semithick, green!50.19607843137255!black, dashed, forget plot]
table {%
840 -1.7
840 35.7
};
\addplot [semithick, green!50.19607843137255!black, dashed, forget plot]
table {%
880 -1.7
880 35.7
};
\addplot [semithick, green!50.19607843137255!black, dashed, forget plot]
table {%
920 -1.7
920 35.7
};
\addplot [semithick, green!50.19607843137255!black, dashed, forget plot]
table {%
960 -1.7
960 35.7
};
\end{axis}

\end{tikzpicture}
        \caption{Available tokens over time for the two policies.}
        \label{fig:tokens}
    \end{subfigure}	
    \caption{\Gls{aoi} dynamics of the \gls{pq} and \gls{qapa} policies for $T_q=40,\mu_b=0.2,\epsilon=0.2$. The \Gls{pq} policy generally has a lower \gls{aoi}, but the \gls{qapa} policy minimizes the \gls{aoi} at the query instants.}
    \label{fig:Sawtooth1}
\end{figure}

We start by exploring the temporal dynamics of the \gls{aoi} process obtained using the \gls{pq} and the \gls{qapa} policies. Recall that \gls{pq} is optimized to achieve a low \gls{aoi} independent of the query process, while \gls{qapa} minimizes the \gls{aoi} at the query times, using cost functions \eqref{eq:cost_AoI} and \eqref{eq:cost_QAoI}, respectively. Fig.~\ref{fig:sawtooth} shows the \gls{aoi} for queries occurring periodically every $T_q=40$ time slots as indicated by the vertical lines, a packet error probability of $\epsilon=0.2$, and a token rate $\mu_b=0.2$. It is seen that the \gls{pq} policy reduces the \gls{aoi} approximately uniformly across time, while the \gls{qapa} policy consistently tries to reduce the \gls{aoi} in the the slots immediately prior to a query, so that the \gls{aoi} is minimized when the query arrives. This is reflected in Fig.~\ref{fig:tokens}, which shows that the \gls{qapa} policy accumulates energy when the next query is far in the future, unlike \gls{pq}. A consequence of this is that the \gls{qapa} policy generally has a slightly higher average \gls{aoi} than the \gls{pq} policy, but the \gls{qaoi} of the \gls{qapa} is significantly lower than that of the \gls{pq} policy.

\begin{figure*}
    \centering
	\begin{subfigure}[b]{.9\columnwidth}
	    \centering
        \begin{tikzpicture}
  \begin{groupplot}[
    group style={group size=1 by 2,
    vertical sep=0.5cm,
    x descriptions at=edge bottom,
    y descriptions at=edge left},
  enlargelimits=false, ymax=50,
  height=0.4\columnwidth,width=0.9\columnwidth,
    xlabel=$t \mod T_q$]
\nextgroupplot[point meta min=0,
point meta max=0.22,
ylabel=AoI (PQ),
    colorbar,
    colorbar style={%
      ymax=0.2,
      ytick={0.0,0.1,0.2},
      /pgf/number format/precision=2,
      /pgf/number format/fixed,
      /pgf/number format/fixed zerofill}]
\addplot [
matrix plot*,
mesh/cols=40,
point meta=explicit,
] table [meta index=2] {Figures/data/aoi_matrix_02.csv};
\nextgroupplot[%
  point meta min=0,
  point meta max=0.8,
  ylabel=AoI (QAPA),
  colorbar,
  colorbar style={%
  ymax=0.75,
    ytick={0.00,0.25,0.50,0.75},
    /pgf/number format/precision=2,
    /pgf/number format/fixed,
    /pgf/number format/fixed zerofill}]
  \addplot [
  matrix plot*,
  mesh/cols=40,
  point meta=explicit,
  ] table [meta index=2] {Figures/data/qaoi_matrix_02.csv};
\end{groupplot}
\end{tikzpicture}
        \caption{\gls{pq} (upper) and \gls{qapa} (lower) density, $\epsilon=0.2$.}
        \label{fig:matrix_02}
    \end{subfigure}	
	\centering
	\begin{subfigure}[b]{.9\columnwidth}
	    \centering
        \input{Figures/CDF,error=0.2,Tq=40,Pb=0.1}       
        \caption{Complementary CDF, $\epsilon=0.2$.}
        \label{fig:CDF_0.2_40_01}
    \end{subfigure}	
	\begin{subfigure}[b]{.9\columnwidth}
	    \centering
        \begin{tikzpicture}
  \begin{groupplot}[
    group style={group size=1 by 2,
    vertical sep=0.5cm,
    x descriptions at=edge bottom,
    y descriptions at=edge left},
  enlargelimits=false, ymax=50,
  height=0.4\columnwidth,width=0.9\columnwidth,
    xlabel=$t \mod T_q$]
\nextgroupplot[point meta min=0,
point meta max=0.22,
ylabel=AoI (PQ),
    colorbar,
    colorbar style={%
      ymax=0.2,
      ytick={0.0,0.1,0.2},
      /pgf/number format/precision=2,
      /pgf/number format/fixed,
      /pgf/number format/fixed zerofill}]
\addplot [
matrix plot*,
mesh/cols=40,
point meta=explicit,
] table [meta index=2] {Figures/data/aoi_matrix_07.csv};
\nextgroupplot[%
  point meta min=0,
  point meta max=0.32,
  ylabel=AoI (QAPA),
  colorbar,
  colorbar style={%
  ymax=0.3,
    ytick={0.00,0.15,0.30},
    /pgf/number format/precision=2,
    /pgf/number format/fixed,
    /pgf/number format/fixed zerofill}]
  \addplot [
  matrix plot*,
  mesh/cols=40,
  point meta=explicit,
  ] table [meta index=2] {Figures/data/qaoi_matrix_07.csv};
\end{groupplot}
\end{tikzpicture}
        \caption{\gls{pq} (upper) and \gls{qapa} (lower) density, $\epsilon=0.7$.}
        \label{fig:matrix_07}
    \end{subfigure}	
    \begin{subfigure}[b]{.9\columnwidth}
	    \centering
        \input{Figures/CDF,error=0.7,Tq=40,Pb=0.1}      
        \caption{Complementary CDF, $\epsilon=0.7$.}
        \label{fig:CDF_0.7_40_01}
    \end{subfigure}
 \caption{AoI distributions and CDFs for \gls{pq} and \gls{qapa} for $T_q=40,\mu_b=0.1$ and $\epsilon=\{0.2,0.7\}$. (\subref{fig:matrix_02}), (\subref{fig:matrix_07}): AoI distribution for the \gls{pq} and \gls{qapa} at various time instances $t\mod T_q$ for $\epsilon=0.2$ and $\epsilon=0.7$, respectively. \Gls{pq} achieves low \gls{aoi} at all times, \gls{qapa} ensures that the \gls{aoi} is low at the query instants, i.e. $t\mod T_q=0$. (\subref{fig:CDF_0.2_40_01}), (\subref{fig:CDF_0.7_40_01}): Complementary CDF of the \gls{aoi} and \gls{qaoi} achieved by the two policies for $\epsilon=0.2$ and $\epsilon=0.7$. Generally, the \gls{qapa} policy has lower \gls{qaoi} but higher \gls{aoi} than the \gls{pq} policy.}
 \label{fig:CDF}
\end{figure*}

The initial observations from Fig.~\ref{fig:Sawtooth1} can be confirmed by the distribution of the \gls{aoi} as a function of the time since the last query, as illustrated in Fig.~\ref{fig:CDF}. Figs.~\ref{fig:CDF}\subref{fig:matrix_02} and \ref{fig:CDF}\subref{fig:matrix_07} show the probability mass function of the \gls{aoi} conditioned on various time instants $t\mod T_q$, while Figs.~\ref{fig:CDF}\subref{fig:CDF_0.2_40_01} and \ref{fig:CDF}\subref{fig:CDF_0.7_40_01} show the \gls{cdf} of the overall \gls{aoi} and \gls{qaoi}. In the scenario with low error probability, $\epsilon=0.2$, the \gls{aoi} distribution of the \gls{pq} policy is uniform across time (upper plot in Fig.~\ref{fig:CDF}\subref{fig:matrix_02}), while the \gls{qapa} policy has an increasing age as time since the query passes, but a far lower age right before and at the query instant, $t\mod T_q=0$ (lower plot in Fig.~\ref{fig:CDF}\subref{fig:matrix_02}). The resulting \gls{cdf} in Fig.~\ref{fig:CDF}\subref{fig:CDF_0.2_40_01} reveals, as expected, that the \gls{aoi} and the \gls{qaoi} are equivalent for the \gls{pq} policy, as the distribution is the same at any time instant. However, for the \gls{qapa} policy, the \gls{qaoi} is significantly lower than the \gls{aoi}, while the \gls{aoi} is often larger than the \gls{pq} policy's. This is due to the fact that the \gls{qaoi} is only measured at the query instants, at which the age of the \gls{qapa} policy is minimized. Due to the energy constraint, this comes at the cost of a generally higher age, causing a higher \gls{aoi} measured at each time instant. Finally, the staircase appearance in the \gls{cdf} is due to the fact that the queries happen periodically. If the queries were arriving at variable (but known in advance) intervals, then the \gls{qapa} would still have lower \gls{qaoi} than the \gls{pq} query, but its \gls{cdf} would have a different shape.

The same observations apply for the the scenario with high error probability, $\epsilon=0.7$, shown in Figs.~\ref{fig:CDF}\subref{fig:matrix_07} and \ref{fig:CDF}\subref{fig:CDF_0.7_40_01}. Although the \gls{aoi} and \gls{qaoi} 
are higher due to the high packet error rate, the applied policies are similar. The gain that the \gls{qapa} policy achieves by clustering its transmissions close to the query instant is clearly reflected in Fig.~\ref{fig:CDF}\subref{fig:matrix_07} where, although there is a significant probability that the packet immediately prior to the query is lost, the \gls{aoi} distribution at $t\mod T_q=0$ is still concentrated close to one.

\begin{figure*}[t!]
	\centering
	\begin{subfigure}[b]{0.67\columnwidth}
	    \centering
        % This file was created by matplotlib2tikz v0.7.5.
\begin{tikzpicture}

\begin{axis}[
width=\fwidth,
legend cell align={left},
legend style={at={(0.05,0.95)}, anchor=north west,draw=white!80.0!black,nodes={scale=0.8, transform shape}},
legend pos=north west,
tick align=outside,
tick pos=left,
x grid style={white!69.01960784313725!black},
xlabel={$\epsilon$},
xmin=0, xmax=0.85,
xtick style={color=black},
y grid style={white!69.01960784313725!black},
ylabel={Average age},
ymin=0, ymax=120,
ytick style={color=black}
]
\addplot [semithick, color=cyan]
table {%
0.0        19.8699492
0.04473684 20.8127233
0.08947368 21.8669814
0.13421053 23.0077106
0.17894737 24.2712446
0.22368421 25.6759375
0.26842105 27.2790731
0.31315789 29.0759439
0.35789474 31.1013566
0.40263158 33.4215718
0.44736842 36.2183511
0.49210526 39.4380351
0.53684211 43.2634694
0.58157895 47.64182  
0.62631579 53.7496918
0.67105263 61.1679276
0.71578947 70.6276637
0.76052632 83.5191501
0.80526316 102.5331052
0.85       133.6091188
};
\addlegendentry{AoI (PQ)}
\addplot [semithick, color=orange_D]
table {%
0.0         19.865892
0.04473684  20.815145
0.08947368  21.864857
0.13421053  23.008087
0.17894737  24.273994
0.22368421  25.684916
0.26842105  27.275813
0.31315789  29.087526
0.35789474  31.100773
0.40263158  33.424841
0.44736842  36.215365
0.49210526  39.455468
0.53684211  43.263105
0.58157895  47.643758
0.62631579  53.751516
0.67105263  61.158975
0.71578947  70.631174
0.76052632  83.526334
0.80526316 102.527568
0.85       133.632507
};
\addlegendentry{QAoI (PQ)}
\addplot [semithick, color=cyan,dashed]
table {%
0.0         19.6825624
0.04473684  20.833293 
0.08947368  21.8614314
0.13421053  22.9687161
0.17894737  24.1905942
0.22368421  25.5818716
0.26842105  27.0722632
0.31315789  28.8187745
0.35789474  30.7436093
0.40263158  33.0205036
0.44736842  35.9221456
0.49210526  39.2230111
0.53684211  42.8310289
0.58157895  47.4521917
0.62631579  53.3564439
0.67105263  60.8938184
0.71578947  70.4082953
0.76052632  83.2513892
0.80526316 102.5787165
0.85       132.7236041
};
\addlegendentry{AoI (QAPA)}
\addplot [semithick, color=orange_D,dashed]
table {%
0.0         15.638859
0.04473684  16.532482
0.08947368  17.560613
0.13421053  18.678701
0.17894737  19.951176
0.22368421  21.410759
0.26842105  22.970561
0.31315789  24.808033
0.35789474  26.840779
0.40263158  29.22148 
0.44736842  32.152901
0.49210526  35.558885
0.53684211  39.189226
0.58157895  43.831456
0.62631579  49.805467
0.67105263  57.353687
0.71578947  66.878055
0.76052632  79.737413
0.80526316  99.099055
0.85       129.2927  
};
\addlegendentry{QAoI (QAPA)}
\end{axis}

\end{tikzpicture}       
        \caption{Average age for $T_q=10,\mu_b=0.05$.}
        \label{fig:avg_10_005}
    \end{subfigure}	
    \begin{subfigure}[b]{0.67\columnwidth}
	    \centering
        % This file was created by matplotlib2tikz v0.7.5.
\begin{tikzpicture}

\begin{axis}[
width=\fwidth,
legend cell align={left},
legend style={draw=white!80.0!black,nodes={scale=0.8, transform shape}},
legend pos=north west,
tick align=outside,
tick pos=left,
x grid style={white!69.01960784313725!black},
xlabel={$\epsilon$},
xmin=0, xmax=0.85,
xtick style={color=black},
y grid style={white!69.01960784313725!black},
ylabel={Average age},
ymin=0, ymax=120,
ytick style={color=black}
]
\addplot [semithick, color=cyan]
table {%
0.0         19.8699492
0.04473684  20.8127233
0.08947368  21.8669814
0.13421053  23.0077106
0.17894737  24.2712446
0.22368421  25.6759375
0.26842105  27.2790731
0.31315789  29.0759439
0.35789474  31.1013566
0.40263158  33.5057886
0.44736842  36.2196292
0.49210526  39.4304671
0.53684211  43.2635248
0.58157895  47.7417574
0.62631579  53.7489735
0.67105263  60.9959821
0.71578947  70.6276735
0.76052632  83.3197858
0.80526316 102.5401967
0.85       133.0962904
};
%\addlegendentry{PQ,AoI}
\addplot [semithick, color=orange_D]
table {%
0.0         19.86259 
0.04473684  20.821014
0.08947368  21.839198
0.13421053  23.015004
0.17894737  24.260326
0.22368421  25.70119 
0.26842105  27.268012
0.31315789  29.075434
0.35789474  31.116994
0.40263158  33.49704 
0.44736842  36.190564
0.49210526  39.440616
0.53684211  43.292618
0.58157895  47.759812
0.62631579  53.771174
0.67105263  61.013702
0.71578947  70.653374
0.76052632  83.316172
0.80526316 102.50396 
0.85       133.078028
};
%\addlegendentry{PQ,QAoI}
\addplot [semithick, color=cyan,dashed]
table {%
0.0         19.8600944
0.04473684  21.3844587
0.08947368  22.5202605
0.13421053  23.6188204
0.17894737  24.8032493
0.22368421  26.1250023
0.26842105  27.5038299
0.31315789  29.2314041
0.35789474  31.1655594
0.40263158  33.501775 
0.44736842  36.2201417
0.49210526  39.3252468
0.53684211  43.2937836
0.58157895  47.7035149
0.62631579  53.3541089
0.67105263  60.8335588
0.71578947  69.8423769
0.76052632  83.8323883
0.80526316 102.4815512
0.85       133.3447164
};
%\addlegendentry{QAPA,AoI}
\addplot [semithick, color=orange_D,dashed]
table {%
0.0         11.82828 
0.04473684  12.737644
0.08947368  13.601688
0.13421053  14.712758
0.17894737  15.93559 
0.22368421  17.309108
0.26842105  18.80642 
0.31315789  20.65133 
0.35789474  22.611306
0.40263158  24.989188
0.44736842  27.888018
0.49210526  31.009054
0.53684211  34.991992
0.58157895  39.417916
0.62631579  45.08037 
0.67105263  52.571026
0.71578947  61.595902
0.76052632  75.64223 
0.80526316  94.339182
0.85       125.269978
};
%\addlegendentry{QAPA,QAoI}
\end{axis}

\end{tikzpicture}       
        \caption{Average age for $T_q=20,\mu_b=0.05$.}
        \label{fig:avg_20_005}
    \end{subfigure}
	\begin{subfigure}[b]{0.67\columnwidth}
	    \centering
        % This file was created by matplotlib2tikz v0.7.5.
\begin{tikzpicture}

\begin{axis}[
width=\fwidth,
legend cell align={left},
legend style={draw=white!80.0!black,nodes={scale=0.8, transform shape}},
legend pos=north west,
tick align=outside,
tick pos=left,
x grid style={white!69.01960784313725!black},
xlabel={$\epsilon$},
xmin=0, xmax=0.85,
xtick style={color=black},
y grid style={white!69.01960784313725!black},
ylabel={Average age},
ymin=0, ymax=120,
ytick style={color=black}
]
\addplot [semithick, color=cyan]
table {%
0.0         19.8699492
0.04473684  20.8127233
0.08947368  21.8669814
0.13421053  23.0077106
0.17894737  24.2712446
0.22368421  25.6759375
0.26842105  27.2790731
0.31315789  29.0759439
0.35789474  31.1013566
0.40263158  33.5053873
0.44736842  36.2196356
0.49210526  39.4304579
0.53684211  43.1252658
0.58157895  47.9117085
0.62631579  53.7531031
0.67105263  60.9958885
0.71578947  70.2331163
0.76052632  83.518984 
0.80526316 102.5399423
0.85       133.0965224
};
%\addlegendentry{PQ,AoI}
\addplot [semithick, color=orange_D]
table {%
0.0         19.884288
0.04473684  20.838828
0.08947368  21.825176
0.13421053  23.011412
0.17894737  24.238616
0.22368421  25.676612
0.26842105  27.263964
0.31315789  29.099864
0.35789474  31.127088
0.40263158  33.449788
0.44736842  36.229856
0.49210526  39.463768
0.53684211  43.118004
0.58157895  47.851484
0.62631579  53.741916
0.67105263  61.00062 
0.71578947  70.25296 
0.76052632  83.520932
0.80526316 102.529628
0.85       133.19336 
};
%\addlegendentry{PQ,QAoI}
\addplot [semithick, color=cyan,dashed]
table {%
0.0         22.7365224
0.04473684  24.7907886
0.08947368  25.8582289
0.13421053  26.7867641
0.17894737  27.9162055
0.22368421  29.0708401
0.26842105  30.3700446
0.31315789  31.8581182
0.35789474  33.6371536
0.40263158  35.8824566
0.44736842  38.4160542
0.49210526  41.3467587
0.53684211  44.986766 
0.58157895  49.4182179
0.62631579  55.2097652
0.67105263  62.2107911
0.71578947  71.661115 
0.76052632  84.4788773
0.80526316 103.2999268
0.85       133.780723 
};
%\addlegendentry{QAPA,AoI}
\addplot [semithick, color=orange_D,dashed]
table {%
0.0          6.628   
0.04473684   7.413212
0.08947368   7.961868
0.13421053   8.881668
0.17894737   9.999032
0.22368421  11.158444
0.26842105  12.513728
0.31315789  14.006704
0.35789474  15.7912  
0.40263158  18.175112
0.44736842  20.860844
0.49210526  23.77146 
0.53684211  27.4052  
0.58157895  31.817092
0.62631579  37.59352 
0.67105263  44.582204
0.71578947  54.036172
0.76052632  66.878032
0.80526316  85.756772
0.85       116.321712
};
%\addlegendentry{QAPA,QAoI}
\end{axis}

\end{tikzpicture}      
        \caption{Average age for $T_q=40,\mu_b=0.05$.}
        \label{fig:avg_40_005}
	\end{subfigure}
		\begin{subfigure}[b]{0.67\columnwidth}
	    \centering
        % This file was created by matplotlib2tikz v0.7.5.
\begin{tikzpicture}

\begin{axis}[
width=\fwidth,
legend cell align={left},
legend style={draw=white!80.0!black,nodes={scale=0.8, transform shape}},
legend pos=north west,
tick align=outside,
tick pos=left,
x grid style={white!69.01960784313725!black},
xlabel={$\epsilon$},
xmin=0, xmax=0.85,
xtick style={color=black},
y grid style={white!69.01960784313725!black},
ylabel={Average age},
ymin=0, ymax=60,
ytick style={color=black}
]
\addplot [semithick, color=cyan]
table {%
0.0         4.8023301
0.04473684  5.0451117
0.08947368  5.3088563
0.13421053  5.6009418
0.17894737  5.9253663
0.22368421  6.2823134
0.26842105  6.6820071
0.31315789  6.9128372
0.35789474  7.4359358
0.40263158  8.0316872
0.44736842  8.746812 
0.49210526  9.5777801
0.53684211 10.5559438
0.58157895 11.7069451
0.62631579 13.1628828
0.67105263 15.0365983
0.71578947 17.4656754
0.76052632 20.7396497
0.80526316 25.6095965
0.85       33.4224137
};
%\addlegendentry{PQ,AoI}
\addplot [semithick, color=orange_D]
table {%
0.0         4.80366 
0.04473684  5.045199
0.08947368  5.31067 
0.13421053  5.60096 
0.17894737  5.92718 
0.22368421  6.284549
0.26842105  6.682474
0.31315789  6.9155  
0.35789474  7.439701
0.40263158  8.033102
0.44736842  8.744375
0.49210526  9.580083
0.53684211 10.559979
0.58157895 11.707259
0.62631579 13.159963
0.67105263 15.032718
0.71578947 17.472138
0.76052632 20.734724
0.80526316 25.610445
0.85       33.424508
};
%\addlegendentry{PQ,QAoI}
\addplot [semithick, color=cyan,dashed]
table {%
0.0         4.7255848
0.04473684  5.3143904
0.08947368  5.575026 
0.13421053  5.8613176
0.17894737  6.1633317
0.22368421  6.4423584
0.26842105  6.8122033
0.31315789  7.2568716
0.35789474  7.7716519
0.40263158  8.3344572
0.44736842  8.9925737
0.49210526  9.7722848
0.53684211 10.7012603
0.58157895 11.8231198
0.62631579 13.1862925
0.67105263 14.9821576
0.71578947 17.2906456
0.76052632 20.6604901
0.80526316 25.3868492
0.85       32.9558125
};
%\addlegendentry{QAPA,AoI}
\addplot [semithick, color=orange_D,dashed]
table {%
0.0         1.98705 
0.04473684  2.183659
0.08947368  2.425125
0.13421053  2.68819 
0.17894737  2.991726
0.22368421  3.269848
0.26842105  3.639138
0.31315789  4.134431
0.35789474  4.729936
0.40263158  5.278386
0.44736842  5.924605
0.49210526  6.700323
0.53684211  7.63449 
0.58157895  8.778743
0.62631579 10.16775 
0.67105263 11.996718
0.71578947 14.413371
0.76052632 17.844664
0.80526316 22.698033
0.85       30.360107
};
%\addlegendentry{QAPA,QAoI}
\end{axis}

\end{tikzpicture}      
        \caption{Average age for $T_q=10,\mu_b=0.2$.}
        \label{fig:avg_10_02}
    \end{subfigure}	
    \begin{subfigure}[b]{0.67\columnwidth}
	    \centering
        % This file was created by matplotlib2tikz v0.7.5.
\begin{tikzpicture}

\begin{axis}[
width=\fwidth,
legend cell align={left},
legend style={draw=white!80.0!black,nodes={scale=0.8, transform shape}},
legend pos=north west,
tick align=outside,
tick pos=left,
x grid style={white!69.01960784313725!black},
xlabel={$\epsilon$},
xmin=0, xmax=0.85,
xtick style={color=black},
y grid style={white!69.01960784313725!black},
ylabel={Average age},
ymin=0, ymax=60,
ytick style={color=black}
]
\addplot [semithick, color=cyan]
table {%
0.0         4.8023301
0.04473684  5.0451117
0.08947368  5.3088563
0.13421053  5.6009418
0.17894737  5.9253663
0.22368421  6.2823134
0.26842105  6.6820071
0.31315789  6.9128372
0.35789474  7.4359358
0.40263158  8.0486783
0.44736842  8.7467469
0.49210526  9.5777886
0.53684211 10.5559748
0.58157895 11.7064646
0.62631579 13.1629766
0.67105263 15.0365867
0.71578947 17.4656174
0.76052632 20.7396449
0.80526316 25.665146 
0.85       33.3779429
};
%\addlegendentry{PQ,AoI}
\addplot [semithick, color=orange_D]
table {%
0.0         4.803944
0.04473684  5.04483 
0.08947368  5.30848 
0.13421053  5.60396 
0.17894737  5.923118
0.22368421  6.29028 
0.26842105  6.683206
0.31315789  6.918386
0.35789474  7.438114
0.40263158  8.05115 
0.44736842  8.742706
0.49210526  9.575244
0.53684211 10.557568
0.58157895 11.709068
0.62631579 13.169146
0.67105263 15.051968
0.71578947 17.456994
0.76052632 20.743886
0.80526316 25.669894
0.85       33.351538
};
%\addlegendentry{PQ,QAoI}
\addplot [semithick, color=cyan,dashed]
table {%
0.0         5.2891717
0.04473684  6.5962881
0.08947368  7.042547 
0.13421053  7.4802906
0.17894737  7.7966848
0.22368421  8.0341599
0.26842105  8.4183899
0.31315789  8.7716018
0.35789474  9.1836905
0.40263158  9.6825347
0.44736842 10.2485682
0.49210526 10.8899405
0.53684211 11.6769297
0.58157895 12.6357055
0.62631579 13.9172623
0.67105263 15.655887 
0.71578947 18.0203711
0.76052632 21.2564154
0.80526316 25.8635784
0.85       33.3541012
};
%\addlegendentry{QAPA,AoI}
\addplot [semithick, color=orange_D,dashed]
table {%
0.0         1.18784 
0.04473684  1.30603 
0.08947368  1.460892
0.13421053  1.615636
0.17894737  1.830252
0.22368421  1.903248
0.26842105  2.13527 
0.31315789  2.400718
0.35789474  2.725272
0.40263158  3.112396
0.44736842  3.591968
0.49210526  4.214922
0.53684211  4.97303 
0.58157895  5.958474
0.62631579  7.270016
0.67105263  9.013626
0.71578947 11.430696
0.76052632 14.719862
0.80526316 19.414892
0.85       27.001714
};
%\addlegendentry{QAPA,QAoI}
\end{axis}

\end{tikzpicture}        
        \caption{Average age for $T_q=20,\mu_b=0.2$.}
        \label{fig:avg_20_02}
    \end{subfigure}
	\begin{subfigure}[b]{0.67\columnwidth}
	    \centering
        % This file was created by matplotlib2tikz v0.7.5.
\begin{tikzpicture}

\begin{axis}[
width=\fwidth,
legend cell align={left},
legend style={draw=white!80.0!black,nodes={scale=0.8, transform shape}},
legend pos=north west,
tick align=outside,
tick pos=left,
x grid style={white!69.01960784313725!black},
xlabel={$\epsilon$},
xmin=0, xmax=0.85,
xtick style={color=black},
y grid style={white!69.01960784313725!black},
ylabel={Average age},
ymin=0, ymax=60,
ytick style={color=black}
]
\addplot [semithick, color=cyan]
table {%
0.0         4.8023301
0.04473684  5.0451117
0.08947368  5.3088563
0.13421053  5.6009418
0.17894737  5.9253663
0.22368421  6.2823134
0.26842105  6.6820071
0.31315789  6.9128372
0.35789474  7.4359358
0.40263158  8.04849  
0.44736842  8.7494401
0.49210526  9.5777908
0.53684211 10.5303448
0.58157895 11.7291211
0.62631579 13.1627806
0.67105263 15.0366377
0.71578947 17.4586421
0.76052632 20.7604627
0.80526316 25.6098263
0.85       33.3315793
};
%\addlegendentry{PQ,AoI}
\addplot [semithick, color=orange_D]
table {%
0.0         4.814736
0.04473684  5.03364 
0.08947368  5.311988
0.13421053  5.610796
0.17894737  5.913616
0.22368421  6.288904
0.26842105  6.681472
0.31315789  6.923644
0.35789474  7.447524
0.40263158  8.048588
0.44736842  8.747956
0.49210526  9.580528
0.53684211 10.508056
0.58157895 11.7474  
0.62631579 13.160536
0.67105263 15.017576
0.71578947 17.45972 
0.76052632 20.751672
0.80526316 25.58422 
0.85       33.330592
};
%\addlegendentry{PQ,QAoI}
\addplot [semithick, color=cyan,dashed]
table {%
0.0         9.8223072
0.04473684 10.3898256
0.08947368 11.0777832
0.13421053 11.4942637
0.17894737 12.0110538
0.22368421 12.24072  
0.26842105 12.7459278
0.31315789 13.3065378
0.35789474 13.7420007
0.40263158 14.278902 
0.44736842 14.7100524
0.49210526 15.3868203
0.53684211 16.0052842
0.58157895 16.7054815
0.62631579 17.6966928
0.67105263 18.9521942
0.71578947 20.7951769
0.76052632 23.3422176
0.80526316 27.5978275
0.85       34.6675732
};
%\addlegendentry{QAPA,AoI}
\addplot [semithick, color=orange_D,dashed]
table {%
0.0         1.00448 
0.04473684  1.084192
0.08947368  1.170084
0.13421053  1.260236
0.17894737  1.370896
0.22368421  1.381008
0.26842105  1.495876
0.31315789  1.607976
0.35789474  1.762076
0.40263158  1.962836
0.44736842  2.20016 
0.49210526  2.535096
0.53684211  2.945212
0.58157895  3.53202 
0.62631579  4.35734 
0.67105263  5.501708
0.71578947  7.252544
0.76052632  9.726752
0.80526316 13.8786  
0.85       20.907668
};
%\addlegendentry{QAPA,QAoI}
\end{axis}

\end{tikzpicture}     
        \caption{Average age for $T_q=40,\mu_b=0.2$.}
        \label{fig:avg_40_02}
	\end{subfigure}
 \caption{Average \gls{aoi} and \gls{qaoi} for the two systems for different values of $T_q$ and $\mu_b$.}
 \label{fig:average_age}
\end{figure*}
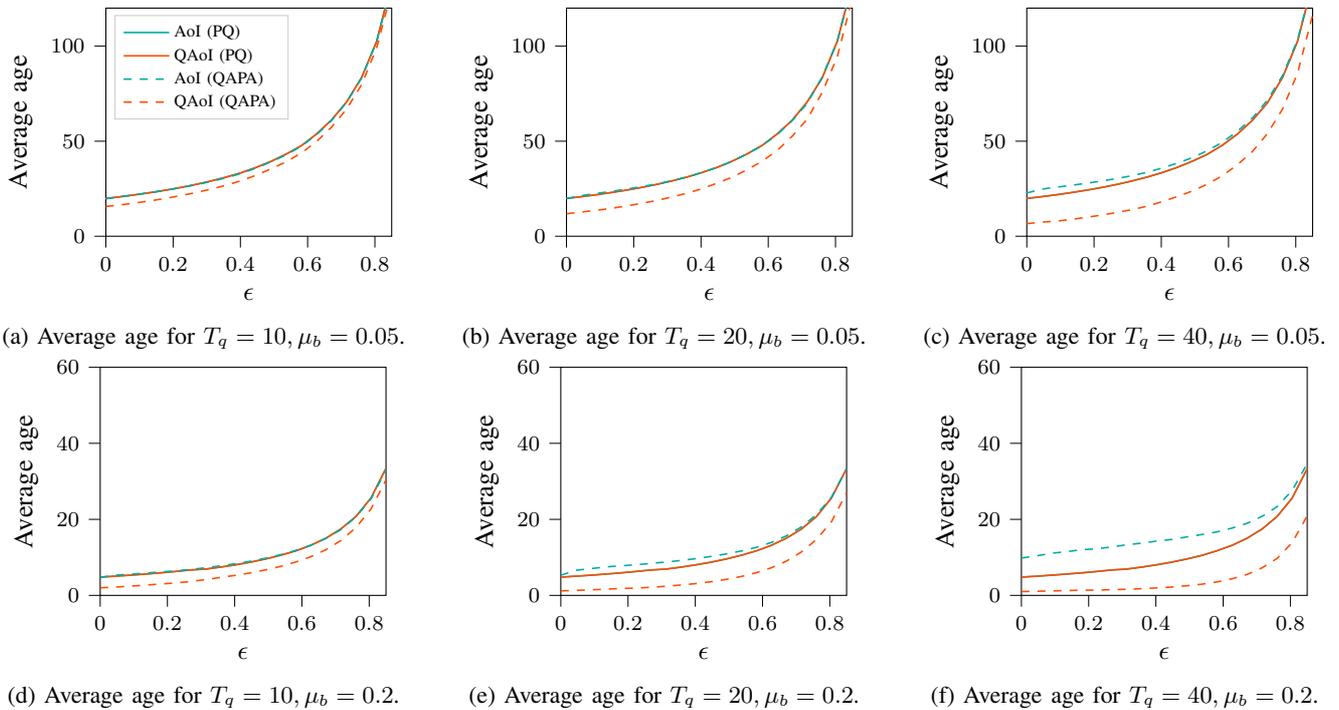

We close the section by studying how the average \gls{aoi} and \gls{qaoi} changes with the packet error probability $\epsilon$ for various choices of the parameters, shown in Fig.~\ref{fig:average_age}. For all cases, the \gls{qapa} policy achieves the lowest \gls{qaoi}, while the \gls{pq} policy achieves the lowest \gls{aoi}. When the query period, $T_q$, is low, the difference between \gls{aoi} and \gls{qaoi} is relatively small, as is the difference between the two policies. Intuitively, this is because the query instants, which are prioritized by the \gls{qapa} policy, are more frequent, making the two problems more similar. If we set $T_q=1$, the two policies would coincide.
As a result, awareness of the query arrival process becomes more important when queries are rare, i.e., when $T_q$ is large: this is clear from the large gap between the average \gls{qaoi} achieved by \gls{qapa} and by \gls{pq} in Fig.~\ref{fig:avg_40_005} and Fig.~\ref{fig:avg_40_02}.
The upper row, Fig.~\ref{fig:average_age}\subref{fig:avg_10_005}-\ref{fig:average_age}\subref{fig:avg_40_005}, shows the results for $\mu_b=0.05$, i.e., when a new token is generated on average every 20 time slots. When $T_q=10$ (\ref{fig:average_age}\subref{fig:avg_10_005}), the token period becomes a limiting factor, and both the \gls{aoi} and \gls{qaoi} are relatively high even for low values of $\epsilon$. In particular, in the error-free case when $\epsilon=0$, the average \gls{qaoi} cannot be lower than $(1+11)/2=6$, which is achieved by transmitting an update prior to every second query. Interestingly, the impact of the energy limit becomes less significant for the \gls{qapa} policy as the time between queries increases: by saving up tokens until right before the query, this policy can significantly reduce the \gls{qaoi}, at the cost of a higher \gls{aoi}. On the other hand, the \gls{pq} policy does not benefit from this increase, as it is oblivious of the query arrival frequency. When tokens are generated faster, at rate $\mu_b=0.2$, as shown in Fig.~\ref{fig:average_age}\subref{fig:avg_10_02}-\ref{fig:average_age}\subref{fig:avg_40_02}, the \gls{aoi} and the \gls{qaoi} are generally lower, since more frequent transmissions are allowed.

\section{Conclusions and future work}\label{sec:concl}
In this paper, we proposed a new metric for measuring information freshness, which we dubbed \gls{qaoi}: unlike standard \gls{aoi} for push-based communication,  this metric can be used for pull-based communication in which the monitoring process is not always listening, but sends queries when it is interested in the information.
With the proposed model and subsequent \gls{mdp} solution, we show the benefit of optimizing the transmission policy using the available knowledge on the query arrival process in terms of \gls{qaoi}. Our results show that the standard \gls{pq} optimization, which minimizes \gls{aoi} at any instant, can be very different from a \gls{qapa} policy that optimizes \gls{qaoi} by concentrating its transmissions right before it expects a new query.

There are several possible avenues of future work we are considering, including a formal derivation of the \gls{qaoi} in simple queuing system, as well as the modeling of more complex query processes with stochastic timing, which would require the sensors to learn the nature of the query arrival process online.

\section*{Acknowledgment}
This work has been in part supported the Danish Council for Independent Research (Grant Nr. 8022-00284B SEMIOTIC).


\end{document}